\documentclass[fleqn,usenatbib]{mnras}

\usepackage[T1]{fontenc}
\usepackage{ae,aecompl}

\usepackage{newtxtext,newtxmath}

\usepackage{graphicx}	

\usepackage{tikz}
\usepackage{pgfplots}
\pgfplotsset{compat=1.14}

\newcommand{\Msun}{\mbox{$M_\odot$}}
\newcommand{\sqdeg}{\mbox{$\mbox{deg}^2$}}

\title[The Counterpart to S190814bv]{Limits on the Electromagnetic Counterpart to S190814bv}

\author[A.~M.~Watson et al.]{%
A.~M.~Watson,$^{1}$\thanks{E-mail: alan@astro.unam.mx (AMW)}
N.~R.~Butler,$^{2}$
W.~H.~Lee,$^{1}$
R.~L.~Becerra,$^{1}$
M.~Pereyra,$^{3}$\newauthor
F.~Angeles,$^{1}$
A.~Farah,$^{1}$
L.~Figueroa,$^{3}$
D.~G\'onzalez-Buitrago,$^{3}$
F.~Quir\'os,$^{3}$\newauthor
J.~Ru\'iz-D\'iaz-Soto,$^{1}$
C.~Tejada de Vargas,$^{3}$
S.~J.~Tinoco,$^{1}$
and
T.~Wolfram$^{2}$
\\
$^1$ Instituto de Astronom{\'\i}a, Universidad Nacional Aut\'onoma de M\'exico, Apartado Postal 70-264, 04510 M\'exico, CDMX, Mexico\\
$^2$ School of Earth and Space Exploration, Arizona State University, Tempe, AZ 85287, USA\\
$^3$ Instituto de Astronom{\'\i}a, Universidad Nacional Aut\'onoma de M\'exico, 22860 Ensenada, BC, Mexico
}

\date{Accepted 2020 January 14. Received 2019 December 31; in original form 2019 October 14}

\pubyear{2020}

\begin{document}
\label{firstpage}
\pagerange{\pageref{firstpage}--\pageref{lastpage}}
\maketitle

\begin{abstract}
We derive limits on any electromagnetic counterpart to the compact binary merger S190814bv, whose parameters are consistent with the merger of a black hole and a neutron star. We present observations with the new wide-field optical imager DDOTI d{and} also consider {\itshape Swift}/BAT observations reported by \cite{25341}. We show that {\itshape Swift}/BAT would  have detected a counterpart with similar properties to a typical on-axis short GRB at the 98 per cent confidence level, whereas our DDOTI observations only rule out such a counterpart at the 27 per cent confidence level. Neither have sufficient sensitivity to rule out an off-axis counterpart like GW 170817. We compare the efficiency of {\itshape Swift}/BAT and DDOTI for future observations, and show that DDOTI is likely to be about twice as efficient as  {\itshape Swift}/BAT for off-axis events up to about 100 Mpc.
\end{abstract}




\section{Introduction}
\label{section:introduction}


Compact binary mergers are exciting probes of stellar evolutionary pathways, the physics of matter at nuclear density, nucleosynthesis, gravitational waves, the birth of black holes, and the large-scale structure of the Universe \citep{1971swng.conf..539W,1974ApJ...192L.145L}. LIGO and Virgo have opened the gravitational wave universe to astronomy \citep{2016PhRvL.116f1102A} and revealed spectacular results on binary black-hole mergers and the detection of the binary neutron-star merger GW170817 both in gravitational waves and in light as an off-axis short gamma-ray burst (SGRB) with a kilonova \citep{2017PhRvL.119p1101A,2017ApJ...848L..12A,2017Sci...358.1556C}. 

The recent merger event S190814bv \citep{25324,25333} is the first to have parameters that are consistent with the merger of a black hole (BH) and a neutron star (NS), a class of event whose viability as a source of gravitational waves and electromagnetic emission had been explored but remained hitherto unobserved \citep{2007NJPh....9...17L,2007PhR...442..166N}. At the time of writing, three other likely BH-NS mergers have been detected \citep{25695,25814,25876}, but these events have much higher false-alarm rates and are either more distant, much less well localized, or both.

LIGO and Virgo provide detections, approximate 3D positions with uncertainties of 10--1000 {\sqdeg}, and basic parameters of the merging and merged objects. However, electromagnetic observations can lead to a much better understanding of the consequences of the merger and its relation to other astrophysical phenomena. A necessary step in this development is the localization of the event to the precision necessary for observations with narrow-field instruments. One option is space-based, wide-field $\gamma$-ray imaging with {\itshape Swift}/BAT.
A second is ground-based, narrow-field optical imaging of individual catalogued galaxies in the detection volume \citep{2016ApJ...820..136G}, which worked spectacularly well in the case of GW 170817 \citep{2017Sci...358.1556C} and which we pursue with our narrow-field telescopes. A final possibility is ground-based, wide-field optical imaging with instruments such as ZTF \citep{2019PASP..131a8002B}, ATLAS \citep{2018PASP..130f4505T}, Pan-STARRS1 \citep{2004AN....325..636H}, DECam \citep{2015AJ....150..150F}, MASTER-Net \citep{2010AdAst2010E..30L}, GOTO \citep{2018cosp...42E2486O}, MeerLICHT \citep{2016SPIE.9906E..64B}, KMTNet \citep{2016JKAS...49...37K}, the TAROT telescopes in the GRANDMA collaboration \citep{2019MNRAS.tmp.2740A}, or our own DDOTI \citep{2016SPIE.9910E..0GW}. In these instruments there are trade-offs between aperture, field, and cost, and DDOTI represents one extreme, having the widest field (69 {\sqdeg}), one of the smallest apertures (28 cm), and one of the lowest costs (about US\$350,000 for hardware). The two ground-based approaches are not completely disjoint. For example, we often pass marginal candidates detected by DDOTI for confirmation with our more sensitive, narrow-field instruments RATIR \citep{2012SPIE.8446E..10B,2012SPIE.8444E..5LW} and COATLI \citep{2016SPIE.9908E..5OW}.

Limits on the electromagnetic behaviour of BH-NS mergers are a vital step in constraining the properties and nature of these events. In this paper we discuss limits on any electromagnetic counterpart to S190814bv provided by observations with {\itshape Swift}/BAT and DDOTI. For DDOTI, the results presented here build on the initial report of \cite{25352}. We then generalize our analysis to consider the efficiencies of the two instruments for localizing the counterparts of future events, and show that DDOTI is likely to be quite efficient especially for off-axis events up to about 100~Mpc.




\section{Observations}
\label{section:observations}

\subsection{LIGO and Virgo Observations}

\cite{25324,25333} reported the compact binary merger event S190814bv detected at $T =$~2019 August 14 21:10:39.013 UTC by both of the LIGO detectors and the Virgo detector. The detection was highly significant, with a false alarm rate of about $10^{-25}$ yr. 


The sky maps were progressively refined. The first was generated at $T + 0.35$~h used BAYESTAR \citep{2016PhRvD..93b4013S}, included data from only the LIGO Livingston and Virgo detectors, and had a 90 per cent area of 772~{\sqdeg}. An update at $T + 1.79$~h added data from the LIGO Hanford detector and reduced the 90 per cent area to 38~{\sqdeg}. A further update at $T+11.94$~h used LALInference \citep{2015PhRvD..91d2003V}, further reduced the 90 per cent area to 23~{\sqdeg}, and gave a luminosity distance of $267 \pm 52$~Mpc ($1\sigma$ uncertainty). This distance corresponds to a redshift of $z = 0.059\pm0.011$.

The LALInference analysis gave a 99.8 probability for a merger with $m_1 \ge 5\Msun$ and $m_2 \le 3\Msun$, a possible NS-BH merger, but negligible probability of material outside the merged object according to equation (4) of \cite{2018PhRvD..98h1501F}. At the time of writing, this is the only public information on the nature of the merger.

\subsection{$\gamma$-ray Observations}

Upper limits on transient $\gamma$-ray emission from any counterpart to S190814bv close to the merger time were given by \cite{25323} for {\itshape INTEGRAL}/SPI-ACS, \cite{25326} for {\itshape Fermi}/GBM, \cite{25335} for {\itshape AGILE}/MCAL, \cite{25341} for {\itshape Swift}/BAT, \cite{25365} for {\itshape Insight}-HXMT/HE, and \cite{25369} for Konus-{\itshape Wind}. 

The most sensitive limits are from {\itshape Swift}/BAT. \cite{25341} report a $5\sigma$ upper limit on the 15--350~keV flux of $1.17\times10^{-7}~\mathrm{erg\,s^{-1}\,cm^{-2}}$ in 1~s assuming a typical SGRB spectrum and a coverage 99.8 per cent of probability in the updated BAYESTAR map. 

\subsection{DDOTI Observations}

\begin{figure}
\includegraphics[width=\columnwidth]{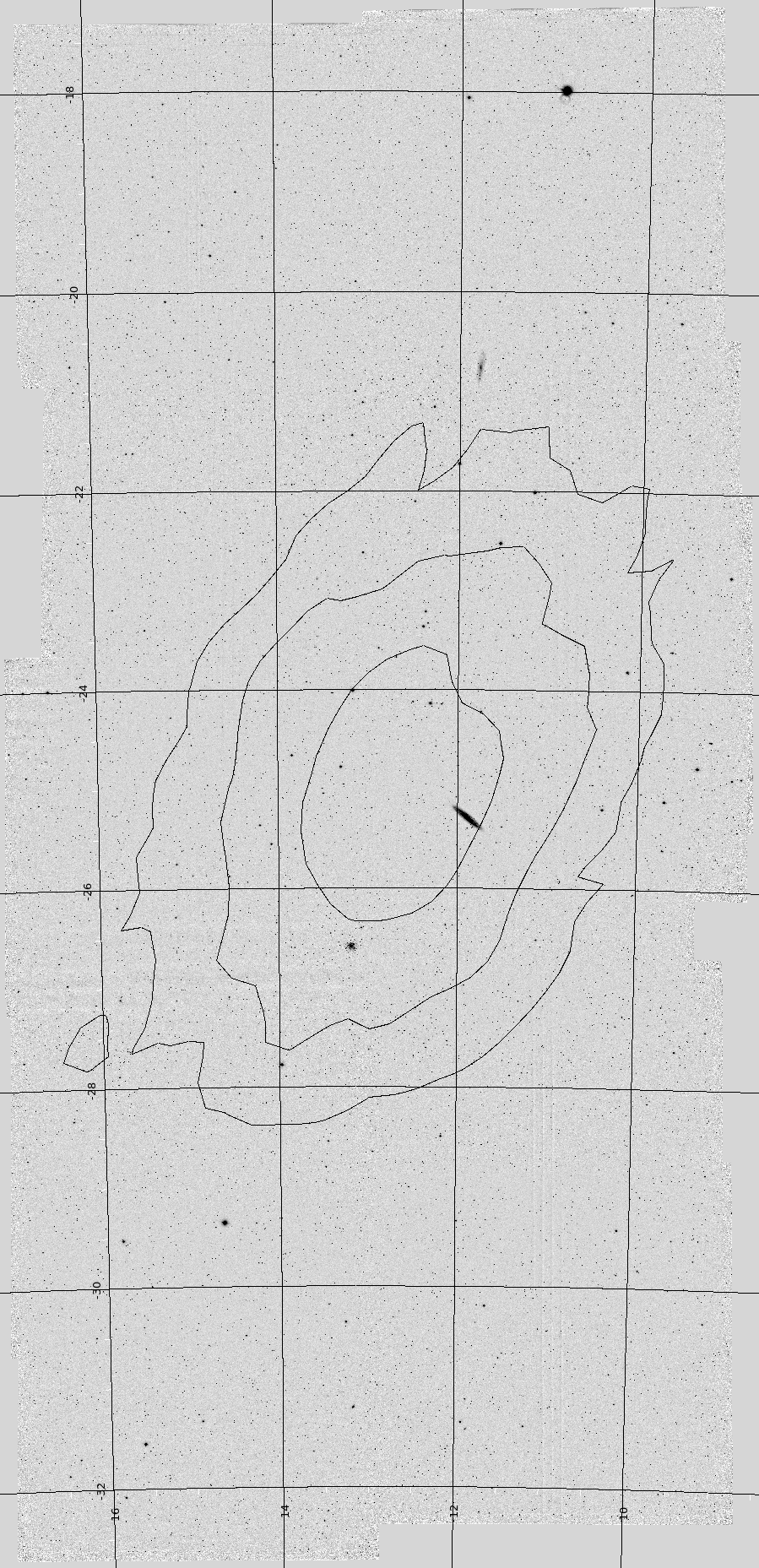}
\caption{The DDOTI image in greyscale. The grid shows J2000 equatorial coordinates with grid-lines every $2~\deg$. The contours show the 2D probability density from LALInference, with contours at 50 per cent, 10 per cent, and 1 per cent of peak.  
The DDOTI observations cover about 100~{\sqdeg}, including the entire dominant region of probability
and the bright stars $\beta$~Cet and $\alpha$~Scl, the nearby galaxies NGC~247 and NGC~253, and the globular cluster NGC~288.  
}
\label{figure:ddoti-image}
\end{figure}

DDOTI (\url{http://ddoti.astroscu.unam.mx/}) is a wide-field, optical, robotic imager located at the Observatorio Astron\'omico Nacional (OAN) on the Sierra de San Pedro M\'artir in Mexico \citep{2016SPIE.9910E..0GW}. It employs an ASTELCO Systems NTM-500 mount with six Celestron RASA 28-cm astrographs each with an unfiltered Finger Lakes Instrumentation ML50100 front-illuminated CCD detector, an adapter of our own design and manufacture that allows static tip-tilt adjustment of the detector, and a modified Starlight Instruments motorized focuser. Each telescope has a field of about $3.4\times3.4~\deg$ with 2.0 arcsec pixels. The individual fields are arranged on the sky in a $2 \times 3$ grid to give a total field of $69~\sqdeg$.

We observed the main region of probability in the updated BAYESTAR map between $T+10.80$~h and $T+14.74$~h (2019 August 15 07:58 and 11:55 UTC). The midpoint of the observations is at $T+12.8$~h. We obtained multiple exposures each of 60~s at different pointings. Because of partial overlapping, the total exposure varies from 1020 to 2820~s. The observing conditions were far from ideal: the target was at airmasses between 2.8 and 1.9, the 99.9 per cent illuminated moon was above the horizon and about $46~{\deg}$ from the target, and about one third of the exposures were taken during astronomical twilight.



Our imaging pipeline subtracts dark images, performs iterative alignment taking into account atmospheric refraction and optical distortions, iteratively estimates and removes the background, resamples to a common pixel grid, performs clipping about the median to remove spurious data and satellite trails, and finally creates a variance-weighted coadded image. It uses {\sc astrometry.net} \citep{2010AJ....139.1782L} for alignment and {\sc swarp} \citep{2010ascl.soft10068B} for stacking.

Our photometry pipeline uses {\sc sextractor} \citep{1996A&AS..117..393B} for source detection and photometry in the natural $w$ AB magnitude system of DDOTI. It uses two aperture diameters, of 3- and 9-times the median stellar FWHM in the image; the difference, for brighter stars and after spatial filtering, is used to estimate the aperture correction, which is then applied locally to the smaller aperture. The calibration is against the APASS DR10 catalog \citep{2018AAS...23222306H} and uses our measured transformation of $w \approx r + 0.23 (g-r)$. The photometry pipeline determines a photometric normalization for each exposure and feeds this back into the imaging pipeline to iteratively correct transparency variations. Our final catalog is produced from the final corrected, coadded image.

The final DDOTI image is shown in Figure~\ref{figure:ddoti-image}. The coverage is about 100~{\sqdeg}, including the entire dominant region in the LALInference probability map, the median $10\sigma$ limiting magnitude is $w_\mathrm{max} = 17.95$, and the mode of the FWHM of the stars with signal-to-noise ratios of greater than 20 is 6.2 arcsec.

Our transients pipeline filters the catalog to identify likely counterparts. It eliminates sources within one FWHM of a USNO-B1 or APASS catalog source, clustered detections (more than 2), near very bright ($R<13$) catalogue stars, near any known minor planets whose positions are supplied by the Minor Planet Center's {\sc mpchecker} service, and whose fluxes in partitions of the data are not consistent with their flux in the final image.

Our final significance level for candidates is set by a number of considerations. One is the desired statistical false alarm probability. Our observations focus on $23~\sqdeg$ and we have a typical FWHM of about 6.2 arcsec, so we have roughly $7.8\times 10^6$ independent elements. For a confidence level of 99 per cent, the appropriate significance is $6\sigma$. A second is the efficiency of our catalog filtering, which in our experience on null fields works well to $10\sigma$ but not to fainter levels. Finally, in sufficiently deep fields we begin to detect large numbers of real but  unrelated astrophysical transients. For S190814bv, the relevant significance level is $10\sigma$, from the efficiency of our filtering in the transients pipeline.

We detect no candidate counterparts at the $10\sigma$ level. Correcting for the estimated Galactic extinction \citep{2011ApJ...737..103S}  of $A_w \approx A_r + 0.23 (A_g - A_r) \approx 0.04$, this corresponds to $w_\mathrm{lim,0} = 17.91$.

\section{Comparison to On-Axis SGRBs}
\label{section:sgrbs}

In this section we consider whether the counterpart of S190814bv might have been an on-axis SGRB. We do this by comparing the limits on emission from any counterpart to the properties of a sample of observed SGRBs. (We note that by ``on-axis'' and ``off-axis'' we refer to the orientation of the GRB jet with respect to the observer and not to the position of the GRB with respect to the instrument boresight.)

The sample of observed SGRBs is the subset of the sample of \cite{2015ApJ...815..102F} with spectroscopic redshifts, supplemented with GRBs 060505 \citep{2007ApJ...662.1129O}, 060614 \citep{2006Natur.444.1053G,2006Natur.444.1050D}, 111117A \citep{2012grb..confE..73S}, 150423A \citep{2015GCN.17755....1M}, and 160821B \citep{2019MNRAS.489.2104T}. We take redshifts and optical observations from these references. We take $\gamma$-ray fluences $S$ and durations $T_{90}$ uniformly from \cite{2016ApJ...829....7L}. The sample contains 34 SGRBs with redshifts, all but one of which (GRB 050709) were discovered by {\itshape Swift}, and includes 8 SGRBs with extended emission ($T_{90} > 2$~s). The redshift range is 0.089 to 2.609.



We note that our sample clearly has uncorrected biases. For example, the SGRB must have had an X-ray localization and must have had either a bright afterglow or a nearby galaxy which allowed the determination of its redshift. We therefore must be careful not to place too much weight on the precise values determined from the subsequent analysis. Nevertheless, the results are indicative.

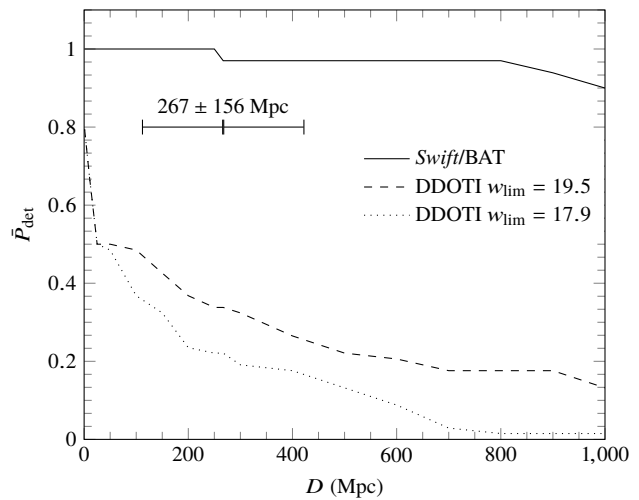
\begin{figure}
\begin{center}
\begin{tikzpicture}
\small
\begin{axis}[
   xmin=0,
   xmax=1000,
   xlabel={$D$ (Mpc)},
   minor x tick num=3,
   ymin=0,
   ymax=1.1,
   minor y tick num=5,
   ylabel={$\bar P_\mathrm{det}$},
   legend entries={{\itshape Swift}/BAT, {DDOTI $w_\mathrm{lim} = 19.5$}, {DDOTI $w_\mathrm{lim} = 17.9$}},
   legend style={
     at={(1,0.7)},
     draw=none,
     cells={anchor=west}
   }
]
\addplot [solid] table [y index=1] {P_det.txt};
\addplot [dashed] table [y index=3] {P_det.txt};
\addplot [dotted] table [y index=2] {P_det.txt};
\draw[|-|] (111,0.8) -- (267,0.8);
\draw[|-|] (267,0.8) -- (423,0.8);
\draw (267,0.85) node {\footnotesize $267 \pm 156$ Mpc};
\end{axis}
\end{tikzpicture}
\end{center}
\caption{The mean probability of detection $\bar P_\mathrm{det}$ for an on-axis SGRB as a function of luminosity distance $D$ within the field of {\itshape Swift}/BAT (solid), DDOTI in good conditions with $w_\mathrm{lim} = 19.5$ (dashed), and DDOTI in poor conditions with $w_\mathrm{lim} = 17.9$ (dotted). For DDOTI, the delay is assumed to be 12.8 h, which is appropriate for our observations of S190814bv and is expected to be typical. Also shown is the $3\sigma$ range for the distance to S190814bv of $267\pm156$~Mpc.}
\label{figure:pdet}
\end{figure}

\subsection{$\gamma$-ray Emission}

The $5\sigma$ limits from {\itshape Swift}/BAT reported by \cite{25341} are our most sensitive limits on $\gamma$-ray emission from any counterpart. (The flight trigger system is complex but effectively works at about $8\sigma$.) Although S190814bv was closer than any of the SGRBs in our sample, the background rate and hence the sensitivity do vary \citep{2005SSRv..120..143B}, so we will perform a brief calculation to confirm our expectation that the non-detection rules out an on-axis SGRB.


\cite{25341} give a $5\sigma$ upper limit on the 15--350~keV flux of $1.17\times10^{-7}~\mathrm{erg\,s^{-1}\,cm^{-2}}$ in 1~s assuming a typical SGRB spectrum. To permit direct comparison to the sample of SGRBs, we convert this to a 15--150~keV flux using their stated power-law spectrum, then assume that $T_{90}$ is approximately equal to the optimal detection interval, and obtain a limit on the 15--150~keV fluence of $6.6\times10^{-8}(T_{90}/1~\mathrm{s})^{1/2}~\mathrm{erg\,cm^{-2}}$.



The observed fluence $S$ and duration $T_{90}$ of a given SGRB scale with redshift $z$ and luminosity distance $D$ such that $SD^2(1+z)^{\Gamma-2}$ and $T_{90}(1+z)^{-1}$ are constant. Here we have assumed that the low-energy $\gamma$-ray photon flux of a SGRB is a power-law $E^{-\Gamma}$ and take $\Gamma = 1.3$ as typical for SGRBs. For the relation between $D$ and $z$, we assume a $\Lambda$CDM universe with cosmological parameters from \cite{2018arXiv180706209P}. Thus, we can determine the fluence and duration that each SGRB in the sample would have as a function of $D$ and compare this to the detection limit. We can further determine the mean probability of detection $\bar P_\mathrm{det}$ for the sample as a function of distance $D$, and we show this in Figure~\ref{figure:pdet}.

With one exception, BAT would easily detect all of the SGRBs in the sample over the distance range to S190814bv. That exception is GRB 150101B, which is expected to be detected only out to about 250 Mpc. That this is the outlier is not surprising; it has the lowest fluence of all SGRBs in our sample and one of the lowest redshifts. That GRB~150101B would be a marginal $5\sigma$ detection at $z=0.059$ whereas it was actually an $8.5\sigma$ detection \citep{2016ApJ...829....7L} at $z=0.134$ seems counterintuitive, but the difference in significance can be understood if the BAT background noise was lower at the time of detection of GRB~150101B.  Assuming a Gaussian distribution for the distance to S190814bv, the mean detection probability is 98.5 per cent. 

Combining the 99.8 per cent coverage reported by \cite{25341} and the 98.5 per cent detection probability determined here, we conclude that if the counterpart to S190814bv had properties similar to those of the observed sample of on-axis SGRBs, there is a probability of about 98.3 per cent that it would have been detected by {\itshape Swift}/BAT. That is, as expected, it is almost certain that any counterpart with properties similar to those of the observed sample of on-axis SGRBs would have been seen by {\itshape Swift}/BAT.

\subsection{Optical Emission}



The observed optical flux density $F_\nu$ of a given SGRB afterglow scales with redshift $z$, luminosity distance $D$, and observed delay $t$ such that $F_\nu D^2t^\alpha(1+z)^{\beta-\alpha}$ is constant. Here we have assumed that the optical flux density behaves as $F_\nu \propto t^{-\alpha}\nu^{-\beta}$ and take  $\alpha = 1$ and $\beta = 0.7$ as typical of SGRB afterglows. Thus, we can determine the optical flux density that each SGRB in the sample would have at 12.8~h as a function of $D$ and compare this to the DDOTI detection limit of $w_\mathrm{lim,0} = 17.91$. Most of our sample have observations in $r$, but some have observations in $g$ or $i$. For our adopted spectrum, we note that  $g-r \approx 0.21$, $r-i\approx 0.15$, and $w-r\approx 0.05$. We can further determine the mean probability of detection $\bar P_\mathrm{det}$ for the sample as a function of distance $D$.

ed{We have only upper limits on the observed flux density for 13 of the 34 SGRBs.} We can define an optimistic case, in which the actual magnitude is just below the upper limit, and a pessimistic case, in which the actual magnitude is so far below the upper limit that the SGRB is never detected by DDOTI. These will bracket the actual detection probability. However, for simplicity we will work with the neutral probability, the average of those in the optimistic and pessimistic cases. This probability is shown as a function of distance in Figure~\ref{figure:pdet}. Assuming a Gaussian distribution for the distance to S190814bv, the mean detection probability is 34 per cent in the optimistic case, 25 per cent in the pessimistic case, and 30 per cent in the neutral case.

The LALInference 2D map contains two main regions of probability, a dominant one at $(\alpha,\delta) =(12.83,-25.24)~\deg$ (J2000) and a secondary one about $12~\deg$ to the southeast. The DDOTI observations, shown in Figure~\ref{figure:ddoti-image}, cover the entire dominant region and contain 89.6 per cent of the probability.

Combining the 89.6 per cent coverage and the 30 per cent detection probability, we conclude that if the counterpart to S190814bv had properties similar to those of the observed sample of on-axis SGRB afterglows, there is a probability of about 27 per cent that it would have been detected by DDOTI.

\section{Comparison to GW~170817}

The compact binary merger GW~170817 was associated with GRB 170817A and the kilonova AT2017gfo \citep{2017Sci...358.1556C,2017ApJ...848L..12A}. AT2017gfo was observed to have $r = 17.14\pm0.08$ and $g-r = 0.18 \pm 0.11$ at about 0.5~d \citep{2017Natur.551...67P} at a distance of about 40~Mpc \citep{2017ApJ...848L..12A}. This magnitude corresponds to $w \approx 17.2$ in the DDOTI system. Our upper limit of $w_\mathrm{lim,0} = 17.9$ suggests that our observations would have detected AT2017gfo only out to about 55~Mpc, which is much closer than the $\pm3\sigma$ derived distance range of 110--420~Mpc for S190814bv \citep{25333}. Our result is similar to that of \cite{25341}, who showed that the $\gamma$-ray emission of a event like GRB 170817A associated with S190814bv would have been detected by {\itshape Swift}/BAT to a distance of only about 70~Mpc. Neither of these observations place a strong constraint on the presence of a counterpart associated with S190814bv similar to that of GW 170817.

\section{Discussion}
\subsection{The Possible Counterpart to S190814bv}

We have shown that the non-detection of $\gamma$-ray emission nearly simultaneous with the merger event, reported by \cite{25341}, conclusively rules out a counterpart similar to a typical on-axis SGRB. This is consistent with our non-detection of an on-axis afterglow with DDOTI, although this non-detection is not so conclusive. 

Nevertheless, the elimination of an on-axis SGRB counterpart is not especially surprising for this event. The prediction of \cite{2018PhRvD..98h1501F}, evaluated with parameters from LALInference \citep{2015PhRvD..91d2003V,25333}, is that the merger left no material beyond the event horizon of the merged object, which suggests a relatively large mass ratio $q$ and relatively low BH spin. A large mass ratio is consistent with the public limit of $q \ge 5/3$. We also note that the publicly available information is also consistent with the secondary being a low-mass $m_2\le 3\Msun$ BH rather than a NS. Finally, even if the counterpart had been similar to a normal SGRB, the probability that it would have been observed on-axis is only 4 per cent for a typical jet angle of $16~{\deg}$ \citep{2015ApJ...815..102F}. 

For this event, neither {\itshape Swift}/BAT not DDOTI have the sensitivity to provide useful constraints on a counterpart similar to AT2017gfo. That is, neither can rule out an off-axis SGRB, a kilonova, or a similarly faint counterpart.

Once the full parameters of the merger are published, including the masses of the components, the highest observed orbital frequency, and the inclination, we will be in a much better position to comment on the nature of any possible counterpart.

\subsection{Future Searches}

LIGO and Virgo typically locate an event to 25--1000~{\sqdeg}. Searches for electromagnetic emissions from gravitational wave events serve two aims. The first is simply detection, but the second is more precise localization, which permits further study. While we enjoy a wide range of $\gamma$-ray detectors in space, the only one that offers realistic prospects for precise localizations is {\itshape Swift}/BAT, since other detectors either have too large positional uncertainties or too small detection possibilities. This leads us to the idea that ground-based searches, which also give localizations, principally compete with or complement {\itshape Swift}/BAT.

The efficiency of an instrument for localization has three parts: the availability (the probability that the instrument can observe), the coverage (the probability of that position of the counterpart is within the field of view of the observations), and the detectability (the probability of detection) or the distance limit for detection.

For {\itshape Swift}/BAT, the availability is essentially 100 per cent. However, its coverage is only about 11 per cent, since the coded field is about 1.4~sr \citep{2005SSRv..120..143B}. S190814bv was a lucky case in which the event occurred in the coded field. {\itshape Swift}/BAT is a superb tool for detecting on-axis emission from GRBs associated with merger counterparts, with a detection probability above 97 per cent to 500~Mpc. This is not at all surprising; {\itshape Swift}/BAT was designed to detect GRBs at cosmological distances and it fulfils these goals very successfully. However,  {\itshape Swift}/BAT is less useful for detecting off-axis emission except in events closer than about 70~Mpc.

DDOTI, as a dedicated ground-based facility at an excellent site,
has an availability of about 79 per cent: 83 per cent for weather  \citep{2009PASP..121..384S} and 95 per cent for system problems (based on our experience since the end of commissioning). Its coverage is about 34 per cent: assuming 75 per cent of events north of $-30~\deg$ declination (for an isotropic distribution of events), 75 per cent of events at least 4 hours from the Sun (again for an isotropic distribution of events), and that DDOTI can cover 50 per cent of the LIGO/Virgo probability map (since DDOTI can cover about 350 {\sqdeg} with an exposure of about 1000 seconds in about two hours of real time, including overheads). The DDOTI detection probability for a given event depends on the observing conditions, the distance to the event, and the delay between the event and the observations. The observations that we report here were in poor conditions and have $w_\mathrm{lim} \approx 17.9$, but we recently observed the HAWC-190917A alert in good conditions (close to the zenith and with the moon below the horizon) and obtained $w_\mathrm{lim} \approx 19.5$ in a similar integration time \citep{25769}. The volume-weighted averages of the neutral detection probability for an on-axis SGRB to 500 Mpc at 12.8~h in poor and good conditions are 17 per cent and 26 per cent, and so we take 20 per cent as an approximate global average. Furthermore, in good conditions the distance to which AT2017gfo could have been detected rises from 55~Mpc to 115~Mpc. For on-axis emission, observing earlier or later significantly increases or decreases the afterglow flux density and so the detection probability, but a delay of 12~h is likely to be typical. For kilonova the temporal evolution should be slower.

In summary, for on-axis SGRB emission out to 500~Mpc, the availability-coverage-detectability product for {\itshape Swift}/BAT is about 11 per cent and for DDOTI is about 5 per cent. For off-axis emission similar to the counterpart to GW 170817, the availability-coverage probability and detection limit for {\itshape Swift}/BAT are 11 per cent and about 70~Mpc and for DDOTI are 27 per cent and 55--115 Mpc. We see that DDOTI is about a factor of two worse than {\itshape Swift}/BAT for on-axis emission but about a factor of two better for off-axis emission.

\section*{Acknowledgements}

We are grateful to Eleonora Troja for sharing her bibliography of SGRBs with redshifts and optical observations and for comments on an earlier draft. We thank the staff of the Observatorio Astron\'omico Nacional. We also thank our industrial partners ASTELCO Systems, Finger Lakes Instrumentation, and Starlight Instruments for their help in deploying the DDOTI hardware. Some of data used in this paper were acquired with the DDOTI instrument of the Observatorio Astron\'omico Nacional on the Sierra de San Pedro M\'artir. DDOTI is funded by CONACyT (LN 260369, LN 271117, and 277901) and the Universidad Nacional Aut\'onoma de M\'exico (CIC and DGAPA/PAPIIT IT102715, IG100414, AG100317, and IN109418) and is operated and maintained by the Observatorio Astron\'omico Nacional and the Instituto de Astronom{\'\i}a of the Universidad Nacional Aut\'onoma de M\'exico.








\bsp	
\label{lastpage}
\end{document}